# Function Interface Models for Hardware Compilation: Types, Signatures, Protocols

Dan R. Ghica

University of Birmingham

`drg@cs.bham.ac.uk`

October 29, 2018

**Abstract**

The problem of synthesis of gate-level descriptions of digital circuits from behavioural specifications written in higher-level programming languages (*hardware compilation*) has been studied for a long time yet a definitive solution has not been forthcoming. The argument of this essay is mainly methodological, bringing a perspective that is informed by recent developments in programming-language theory. We argue that one of the major obstacles in the way of hardware compilation becoming a useful and mature technology is the lack of a well defined *function interface model*, i.e. a canonical way in which functions communicate with arguments. We discuss the consequences of this problem and propose a solution based on new developments in programming language theory. We conclude by presenting a prototype implementation and some examples illustrating our principles.

## 1 Introduction

The problem of hardware compilation turned out to be surprisingly difficult. Although the pioneering work of van Berkel [31], Page, Luk and the Oxford hardware compilation group [23, 18, 13] yielded promising initial results, more than a decade later this technology has yet to enter the mainstream of digital design. Several C-to-gates hardware compilers, such as Roccc [7], Cash [6] and HandelC [8] have been developed, but their take-up was limited. Recently, Mentor has introduced its own C-to-gates compiler, Catapult-C[1] but it is too early to evaluate its impact. In this paper, when we refer to existing hardware compilers we mean these above. We will not refer to design flows based on SystemC[2] or CoWare[3], hardware compilers based on process calculi (e.g. [33, 30]), or higher-order structural languages such as Lava [5]; these are interesting and useful, but conceptually different ways of approaching VLSI design.

---

[1] http://www.mentor.com/products/c-based_design/catapult_c_synthesis
[2] http://www.systemc.org
[3] http://www.coware.com

So, why did hardware compilers not meet expectations?

Part of the answer has to do with performance, as higher-level behavioural design is unlikely to be as efficient as structural or low-level, hand-crafted designs. However, the economics of reconfigurable architectures such as field-programmable gate arrays (FPGAs) or other complex programmable logic devices (CPLDs) are such that the design costs can often become the overriding concern, allowing for weaker performance as a trade-off, especially since their competition is often not custom-design application-specific circuits (ASIC) but (embedded) software. This is a well known consideration, so hardware compilers often target reconfigurable architectures.

Another performance-related issue is that of concurrency; the performance advantage of hardware comes from its potential for massive parallelism, rather than high clock rates. Conventional programming languages that can serve as candidates for compilation into hardware (C, Java, etc.) have either no built-in support for it or offer unsuitable concurrency models such as threads or processes. This is a serious dilemma, because a concurrency model needs to be both low-level enough to reflect the underlying capabilities of the platform ("*close-to-metal*") and thus give predictable performance, but also high-level enough to unburden the programmer from detailed management and book-keeping of resources. An example of a successful such trade-off is Nvidia's CUDA platform[4]. However, the importance of this argument must not be overstated. A recent study showed that a large class of non-trivial structural designs can be re-created automatically by a hardware compiler from behavioural specifications, using standard optimisation techniques such a parallelisation [27].

While the arguments above carry some strength and reflect the conventional wisdom on the topic, we will argue that we must look deeper in order to discover the key problems that beset hardware compilation, and to find ways to address them. In this essay we propose that one such key problem is the lack of canonical function interface models (FIM). We will briefly examine their traditional role in software compilation and operating systems, and explain the possible reasons for their absence in hardware compilers. We also propose a simple but non-trivial FIM, illustrating it with several typical examples and explain how such FIMs can be canonically designed.

§

In modern programming-language theory the notion of *type* is paramount. Beyond just classifying *data* in various categories such as integers, floating-point, strings, etc., types fill a much more fundamental role. It is difficult to give a better summary that does justice to the importance of types than Robert Harper's[5]:

---

[4] http://www.nvidia.com/cuda
[5] http://www.cs.cmu.edu/~rwh/research.htm

> *Over the last two decades type theory has emerged as the central organizing framework for the design and implementation of programming languages. Type theory (the study of type systems) provides the theoretical foundation for safe component integration. In the words of John Reynolds, "a type system is a syntactic discipline for enforcing levels of abstraction". By* syntactic *we mean that the type system is* part of the program, *rather than purely in the mind of the implementer. By* discipline *we mean that* type restrictions are enforced; *ill-typed combinations are ruled out by the type system. By* levels of abstraction *we mean the clean separation between conceptually distinct data objects that may, in a particular program or compiler, have the same or similar representations.*
>
> *Implicit in this definition is an important principle:* the power of a type system lies as much in what it precludes as what it allows. *The most powerful type system of all is the one that rules out all programs. However, such a type system, while surely preventing safety violations, is hardly very useful. The goal of type system design is to* increase expressiveness *by admitting useful programs,* but without compromising safety.

We will see in the following how type systems are indeed essential in hardware compilation of programming languages because of the special challenge of "safe component integration" in hardware, as opposed to software.

## 2 Function interface models

Functions and related concepts (procedures, methods, subroutines, etc.) are the main mechanism for implementing *abstraction* in a programming language. The importance of using functional abstractions lies at the core of any good programming methodology and its benefits have long been established. Functions play a fundamental role in the operation of a conventional stored-program computer and they were in fact a feature of the first such computer, the EDSAC [34]. Except for very early models such as the HP 2100, microprocessors always supported function call in their instruction set directly (e.g. Intel's x86) or at least provided instructions for stack management, meant mainly to implement function calls (RISC architectures).

All three initial major modern programming languages (FORTRAN, LISP, COBOL) provided support for functions, albeit sometimes in a clumsy or limited way. Of the three, the most advanced support for functions is found in LISP. Not only does it support *higher-order* functions (functions that take functions as arguments) but it also introduced the new concept of a *Foreign Function Interface*: a way for a program written in LISP to call functions written in another programming language. This idea was adopted by all subsequent programming languages that had mature compilers, under one name or another[6]. A special and privileged position is that of the C programming language; because of the close relationship between C and the operating system its own calling convention is usually implemented directly by the OS and is called the *Application binary interface* (ABI).

---

[6] In JAVA it is called the *Java Native Interface* (JNI).

One of the decisions of the C and Unix designers with the farthest-reaching consequences was to make the details of the calling convention public [16]. The positive effects of this decision cannot be overstated, as it massively improved the interoperability of applications by opening the OS functionality to other programming languages and by facilitating support for stand-alone language-independent libraries.

In this paper, to avoid ambiguity, we introduce the terminology of *function interface model* (FIM) to encompass the closely-related issues of the FFI and ABI.

## 2.1 FIMs and hardware compilation

It is taken as a given that stored-program computers must offer well-defined FIMs. It is inconceivable to design a modern operating system or compiler if this fundamental requirement is not met. On the other hand, in the world of hardware the concept of a FIM simply did not arise. The net-lists (boxes and wires) that are the underlying computational models of hardware languages are not structured in a way that suggests any obvious native FIM.

The abstraction mechanism common in hardware languages such as Verilog or VHDL, the *module*[7], is a form of placing a design inside a conceptual box with a specified interface. This does serve the purpose of hiding the implementation details of the module from its user, which is one of the main advantages of abstraction. However, a module is a less powerful than functional abstraction as found in programming languages in several significant ways:

1. Modules are always *first order*. A module can only take data as input and not other modules; in contrast, functions can take other functions as input.

2. Modules must be explicitly instantiated. The hardware designer must manually keep track of the names of each module occurrence and its ports, which must be connected explicitly by wires to other elements of the design. In contrast, the run-time management of various instances of functions is managed by the OS (using activation records) in a way that is transparent to the programmer.

3. Sharing a module from various points in a design is not supported by hardware design languages. The designer must design ad hoc circuitry, such as multiplexers or de-multiplexers and ad hoc protocols to achieve this. The lack of language support for sharing makes this a delicate and error-prone task.

---

[7]Not to be confused with "modules" as encountered in programming languages such as ML.

These limitations may seem inconsequential, and in a certain sense they are so. They are not limitations on the expressiveness of hardware design languages or the performance of the circuits synthesised. However, the impact of these limitations becomes more serious as designs become more complex. For the design of, say, an adder the limitations above are irrelevant. But if the design is an implementation of a complex algorithm that uses many instances of the same kind of modules, interacting in non-trivial ways the burden of micro-managing modules and ports and the inability to use generic algorithms such as map-reduce, which are inherently higher-order, can become extremely taxing. The burden of managing sharing is also substantial, and the conventional solutions to this problem, such as bus or network-on-chip (NOC) architectures, are complex, heavy-duty and not provided with language-level support.

Generally, existent hardware compilers are not built around well-defined FIMs. What would be the unthinkable in the world of software compilation is the norm in the world of hardware compilation. But this is not entirely surprising, considering the fact that hardware does not have a "native" FIM to offer. Its absence damages the usability and the performance of the compilers, imposes restrictions on interoperability and prevents library support.

Let us briefly consider how functions are managed in the hardware compilers mentioned earlier.

**Roccc** only supports functions via *inlining*, i.e. the textual substitution of the body of the function at the point of call. This on-the-cheap solution to the problem of functional abstraction has a serious problem: it makes sharing of resources impossible. Inlining is a compiler optimisation that targeted to a conventional computer is sometimes justified, trading off better execution time for increased program size. However, from a hardware compilation perspective it is inefficient as it amounts to the re-instantiating the module implementing the function every time it is called. The example that illustrates the failure of this approach best is compilation of floating-point programs: the circuits implementing floating-point operations are expensive and the requirement to re-instantiate multiplication or division every time they are used cannot be satisfied. The hardware-compilation of floating-point programs is impossible if these elementary circuits cannot be shared in the program.

**Cash** proposes a token-based mechanism that allows the implementation of a function to be shared, therefore avoiding the need for re-instantiation. The token-based mechanism is powerful enough to support recursive functions, which is expressive but inefficient. This is a promising approach, and the token-based mechanism represents a genuine FIM. The problem with this approach, inspired by tagged-token data-flow [3], is that it is unnecessarily heavy-duty. The management of the tokens require sophisticated circuitry and access to an external RAM for each function call. This is likely to be expensive in both time and foot-print, and frequent RAM access represents a bottleneck for concurrency. A beta version of the compiler has been announced but is not available yet, so we can only speculate regarding the performance penalty that must be paid in overheads for this token-based mechanism.

**Handel C** seems to provide static support for functions. The details of the FIM are not disclosed by CELOXICA but evidence can be seen in the generated HDL. Unfortunately, the simultaneous support for sharing and concurrency in the absence of a type system that would control such behaviour leads to some surprising race conditions. For example the execution of the concurrent statement `par {x=f(1); y=f(0);}` always stores the same values in registers `x,y`, regardless of what function `f()` computes. The reason is that the simple-minded run-time sharing of the implementation for `f()` causes a race condition on its shared inputs. Having a race condition even in the absence of side-effects is too surprising for a conventional programmer and it must be considered a serious defect in the design of the language. It is not a defect in the implementation of the language because the HANDEL C manual warns that such race conditions may occur. But, concurrent programming, difficult enough as is, becomes almost impossible when confronted with such recondite sources of racing behaviour.

**Catapult C** and a score of other commercial compilers are difficult to asses. The system is a closed one, and Mentor does not publish the details of the support it provides for functions in its technical literature[8]. Even if such languages do have internally a well-defined FIM the fact that it is undisclosed still represents a significant barrier to interoperability and run-time support.

The closest correspondence to a FIM in hardware is the concept of *bus*, a mechanism for establishing *logical* connections between several components using the same set of wires. High-performance bus architectures such as ARM's AMBA[9] or IBM's CORECONNECT[10] are marketed as especially relevant for "*systems-on-chip*" (SoC) designs, as a way of managing complexity and reduce design cost and time. The aims of using a bus architecture are similar to those of using FIMs, but the techniques involved are substantially different. Whereas a bus is a *global* mechanism of *communication* between components, a FIM is a *local* mechanism for providing access to a component. Therefore, bus protocols are complex and layered, more similar to network or distributed-communication protocols. They can handle transmission delay, transmission errors and transactions. In comparison, the FIMs assume perfect, instantaneous (synchronous) communication between components. As a result, interfacing components using a bus requires high overhead and is potential bottleneck for parallelism, whereas managing the FIMs requires typically very low overhead and has some support for parallel access. Buses work best at the architectural level, to connect large components over long distances (e.g. across the boundaries of a chip or of a board) whereas FIMs are meant to handle local connections, with low overhead, low latency (at most one clock cycle in our prototype compiler) and high throughput.

---

[8] http://www.mentor.com/products/esl/techpubs/
[9] http://www.arm.com/products/solutions/AMBAHomePage.html
[10] http://www-306.ibm.com/chips/products/coreconnect/

A further elaboration of the concept of bus is the so-called "*network on chip*" architecture, which makes communication between components even more network-like. For complex applications such as SOC it can be a notable improvement in functionality over the simpler bus technology, but at a correspondingly increased cost [17].

## 3 The design of a hardware FIM

The previous section looked at the state-of-play in hardware compilation. Hopefully the reader will have accepted the basic point that well-defined FIMs are important and useful and it is next to impossible to imagine a mature compiler without them. From here on we will propose a way to derive hardware-friendly FIMs in a canonical way using some recent developments from programming language theory.

It is a well-known fact that relying on the primitive concepts of *channel*, *event* and *communication* makes process calculi good intermediate abstractions for hardware, and refining processes into hardware-level representation has been extensively studied [32]. On the other hand, there has been significant research work regarding the encoding of (prototypical) functional programming languages into process calculi [19].

Game Semantics [1, 15] is a process-calculus-like model for programming languages introduced in the mid-90s, which proved to be extremely successful at giving precise interpretations for a variety of languages, thus solving long-standing open problems in the theory of programming languages. Process calculi are versatile but have little structure, whereas game semantics encapsulates the right mathematical structure needed to interpret programming languages. Like process calculi, it is event-oriented and can be refined into hardware.

Therefore we propose the following methodological principle:

> **Principle 1** *Hardware compilation of programming languages via game semantics is a natural approach based on solid foundational results.*

In the design of the FIM we will consider the following key topics: the static aspects of the FIM, the dynamic aspects of the FIM, and language-level support for FIM compliance. To give focus to our presentation we will use a particular language as a case-study, with the following defining features: a functional fragment based on the affine lambda calculus, combined with the simple imperative language (locally-scoped block variables, iteration, branching), with boolean variables only[11]. This language is a simplified version of *Syntactic Control of Interference* (SCI), a language which has been studied extensively [24, 25, 21, 20].

---

[11]Adding other finite data-types such as integers or floating point is conceptually straightforward, but the technical complications would detract from the main point.

The primitive types of the language are commands, memory cells and (boolean) expressions: $\sigma ::= \mathsf{com} \mid \mathsf{cell} \mid \mathsf{exp}$. Additionally, the language contains function types and products:
$$\theta ::= \sigma \mid \theta \times \theta' \mid \theta \to \theta.$$
What is peculiar about the types above is that pairs of terms *may* share identifiers but functions *may not* share identifiers with their arguments. Terms have types, described by typing judgments of the form $\Gamma \vdash M : \theta$, where $\Gamma = x_1 : \theta_1, \ldots x_n : \theta_n$ is a variable type assignment, $M$ is a term and $\theta$ the type of the term.

$$\frac{}{x : \theta \vdash x : \theta} \text{ Identity}$$

$$\frac{\Gamma \vdash M : \theta}{\Gamma, x : \theta' \vdash M : \theta} \text{ Weakening}$$

$$\frac{\Gamma, x : \theta' \vdash M : \theta}{\Gamma \vdash \lambda x.M : \theta' \to \theta} \to \text{Introduction}$$

$$\frac{\Gamma \vdash F : \theta' \to \theta \quad \Delta \vdash M : \theta'}{\Gamma, \Delta \vdash FM : \theta} \to \text{Elimination}$$

$$\frac{\Gamma \vdash M : \theta' \quad \Gamma \vdash N : \theta}{\Gamma \vdash \langle M, N \rangle : \theta' \times \theta} \times \text{Introduction}$$

The language contains the standard constructs for structured state manipulation and control. It is convenient to present these constructs in a functional form:

| | |
|---:|---:|
| $1 : \mathsf{exp}$ | constant |
| $0 : \mathsf{exp}$ | constant |
| $\mathsf{skip} : \mathsf{com}$ | no-op |
| $\mathsf{asg} : \mathsf{cell} \times \mathsf{exp} \to \mathsf{com}$ | assignment |
| $\mathsf{der} : \mathsf{cell} \to \mathsf{exp}$ | dereferencing |
| $\mathsf{seq} : \mathsf{com} \times \mathsf{com} \to \mathsf{com}$ | sequencing |
| $\mathsf{par} : \mathsf{com} \to \mathsf{com} \to \mathsf{com}$ | parallel execution |
| $\mathsf{op} : \mathsf{exp} \times \mathsf{exp} \to \mathsf{exp}$ | logical operations |
| $\mathsf{if} : \mathsf{exp} \times \mathsf{com} \times \mathsf{com} \to \mathsf{com}$ | branching |
| $\mathsf{while} : \mathsf{exp} \times \mathsf{com} \to \mathsf{com}$ | iteration |
| $\mathsf{newvar} : (\mathsf{cell} \to \mathsf{com}) \to \mathsf{com}$ | local variable. |

Product has syntactic precedence over arrow, which associates to the right.

For example, a term with conventional syntax, such as

```
integer x;
x = 0;
if (x<>y) x=!x;
```

would be written as

$\mathsf{newvar}(\lambda \mathtt{x}.$

$$\texttt{seq}(\texttt{asg}(\texttt{x}, 0),$$
$$\texttt{if}(\texttt{neq}, \texttt{asg}(\texttt{x}, \texttt{not x}), \texttt{skip}))).$$

The interesting structural feature of the type system SCI is allowing sharing of identifiers (contraction) in product-formation but disallowing it in function application. Reynolds, the inventor of SCI, was interested in this rule to eliminate covert interference between terms that ostensibly do not share identifiers, hence the name. This type system facilitates reasoning about program correctness and it has eventually led to the development of bunched [22] and separation logic [26].

This restriction can be exploited in several ways, as noticed in the different type signature on sequential (uncurried) versus parallel (curried) composition. This makes terms such as $\lambda x.x; x$ are legal, while $\lambda x.x \,||\, x$ is illegal. Another consequence of this restriction is that nested function application as in $\lambda f \lambda x.f(f(x))$ is also banned. We note that the type-system SCI is an instance of the type-system *Syntactic Control of Concurrency* (SCC), which has been studied using game semantics [12][12].

### 3.1 Static interfaces for FIM in hardware

As one would perhaps expect, there will be a strong connection between the interface of a hardware module and the signature of the function or, more generally, the term, it implements.

The game-semantic model interprets types as so-called *arenas*, which are structured sets of basic actions called *moves*.

**Definition 1** *An* arena *$A$ is a triple $\langle M_A, \lambda_A, \vdash_A \rangle$ where*

- *$M_A$ is a set of* moves,

- *$\lambda_A : M_A \to \{O, P\} \times \{Q, A\}$ is a function determining for each $m \in M_A$ whether it is an* Opponent *or a* Proponent *move, and a* question *or an* answer.

- *$\vdash_A$ is a binary relation on $M_A$, called* enabling, *satisfying*

    - *if $m \vdash_A n$ for no $m$ then $\lambda_A(n) = (O, Q)$,*
    - *if $m \vdash_A n$ then $\lambda_A^{OP}(m) \neq \lambda_A^{OP}(n)$,*
    - *if $m \vdash_A n$ then $\lambda_A^{QA}(m) = Q$.*

We write $\lambda_A^{OP}, \lambda_A^{QA}$ for the composite of $\lambda_A$ with respectively the first and second projections. If $m \vdash_A n$ we say that *m enables n*. We shall write $I_A$ for the set of all moves of $A$ which have no enabler; such moves are called *initial*. Note that an initial move must be an Opponent question.

The arenas for the basic types are as follows:

---

[12]SCI is SCC with all concurrency bounds set to the unit value.

- $[\![\mathtt{com}]\!] = \langle \{q, a\}, \{q \mapsto (O, Q), a \mapsto (P, A)\}, \{(q, a)\} \rangle$.

  This tells us that a command has two observable actions, corresponding to a request for execution and an acknowledgement that the execution completed successfully.

- $[\![\mathtt{exp}]\!] = \langle \{q, t, f\}, \{q \mapsto (O, Q), t \mapsto (P, A), f \mapsto (P, A)\}, \{(q, t), (q, f)\} \rangle$.

  A boolean expression has three observable actions, a request for execution and two possible outcomes.

- $[\![\mathtt{cell}]\!] = \langle \{q, t, f, wt, wf, a\}, \{q \mapsto (O, Q), t \mapsto (P, A), f \mapsto (P, A), wt \mapsto (O, Q), wf \mapsto (O, Q), a \mapsto (P, A)\}, \{(q, t), (q, f), (wt, a), (wf, a)\} \rangle$.

  A memory cell has three actions for reading, the same as in the case of a boolean expression, and three actions for writing: requests to write true or write false and an acknowledgement.

The enabling relation reflects, intuitively, the basic causal connection between moves: for a base type, an answer must be 'enabled' by a question.

Composite types are interpreted by composite arenas constructed as follows:

- $M_{A \times B} = \langle M_A + M_B, [\lambda_A, \lambda_B], \vdash_A + \vdash_B \rangle$.

  The product arena is simply the structure-preserving disjoint union of the two arenas.

- $M_{A \to B} = \langle M_A + M_B, [\overline{\lambda}_A, \lambda_B], \vdash_A + \vdash_B + I_B \times I_A \rangle$, where $\overline{\lambda} m = (O, x)$ if and only if $\lambda m = (P, x)$.

  The arrow arena is similar to the product arena but it involves the switching of polarities Proponent-Opponent for the arena in the contra-variant position, and extending the enabling relation so that initial moves of the argument are justified by the initial moves of the function.

  The computational intuition for reversing polarities is that, in the case of an argument the "inputs" become "outputs" and vice-versa. The extension of the enabling relation is motivated by establishing a causal connection between computations executed by the argument and those executed by the main body of the function.

For example, the arena for type $\mathtt{com} \to \mathtt{com}$ is:

- $M = \{q1, q2, a1, a2\}$
- $\lambda = \{q1 \mapsto OQ, q2 \mapsto OQ, a1 \mapsto PA, a2 \mapsto PA\}$
- $\vdash = \{(q1, q2), (q1, a1), (q2, a2)\}$.

In the definition of the set of moves, $M_{\mathtt{com} \to \mathtt{com}} = M_{\mathtt{com}} + M_{\mathtt{com}}$ we defined the two injections associated to the disjoint sum as $\mathrm{in}_i(x) = xi$.

There is a natural and elegant connection between an arena and an interface:

**Principle 2** *There is an immediate connection between the following game-semantic and hardware concepts:*

- *An* arena *corresponds to an* interface.
- *A* move *corresponds to a* port *in the interface*.
- *A* Proponent *move is an* output *port*.
- *An* Opponent *move is an* input *port*.
- *A* question *is a* request.
- *An* answer *is an* acknowledgment.

A term of type $\Gamma \vdash M : T$ is interpreted in arena $[\![\Gamma]\!] \to [\![T]\!]$. This means that this term will be compiled into a circuit with interface defined by its arena. For example, for a term $\mathtt{x : com} \vdash \mathtt{M : com}$, the interface will be given by the arena $\mathtt{com} \to \mathtt{com}$ discussed above:

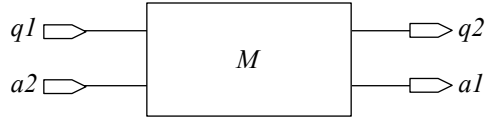

What is not obvious in the conventional interface depiction above is the enabling relation, which plays an essential role in the definition of the semantics and of the FIM. In fact the concept of enabling does not have any obvious hardware equivalent.

## 4 The sequentially reentrant access protocol

In the previous section we gave a method for defining an interface for any type as a structured set of ports. In this section we give a protocol which governs access to these interfaces, rules which will be formulated in an assume-guarantee way: *assuming* the environment is well behaved in the way it provides input actions, a circuit must *guarantee* that its outputs will behave in a certain way. This protocol will be derived directly from the game-semantic model.

Let us start again with some basic game-semantic definitions.

A *justified sequence* in arena $A$ is a finite sequence of moves of $A$ so that the first move is initial and each subsequent occurrence of a move $n$ must have a uniquely associated earlier occurrence of a move $m$ such that $m \vdash_A n$. We say that $n$ is (explicitly) justified by $m$ or, when $n$ is an answer, that $n$ answers $m$.

If a question does not have an answer in a justified sequence, we say that it is *pending* in that sequence. In what follows we use the letters $q$ and $a$ to refer to question- and answer-moves respectively, $m$ will be used for arbitrary moves and $m_A$ will be a move from $M_A$.

A justified sequence is a *legal move* if it satisfies:

**Fork** : In any prefix $s' = \cdots\ q\ \overset{\frown}{\cdots}\ m$ of $s$, the question $q$ must be pending before $m$ is played.

**Wait** : In any prefix $s' = \cdots\ q\ \overleftarrow{\cdots\ a}\ $ of $s$, all questions justified by $q$ must be answered.

**Serial** : In any prefix $s' = \cdots q \cdots q$ the first occurrence of $q$ cannot be pending when the second one is played.

The simplest sequences of moves that *violate* Fork, Wait and Serial respectively are:

**Fork** : $q1 \overleftarrow{\quad a1\quad} m$

**Wait** : $q1 \overleftarrow{\quad q2\quad} a1$

**Serial** : $q1 \overleftarrow{\quad q2\quad} q2$

For an arena $A$, the set of all plays satisfying these conditions is $P_A$, its set of (legal) plays.

The intuition behind these game rules is intimately connected to the language SCI. We can think of a question as the start of a process and an answer as its termination. The FORK rule says essentially that only a "live" process can spawn children; JOIN stipulates that a process can only terminate if all its children processes have terminated. This is essentially the static nature of concurrency in our language, where the only concurrent construct is `par`. The final rule, SERIAL, restricts any process to only one active instance at any given time, which is a consequence of our banning of sharing in concurrent contexts; in the context of hardware it means that each module may receive a new initial request only when the previous initial request has been acknowledged.

The consequence of the SERIAL rule is the following:

**Theorem 2 ([12])** *The representation of the game model for any term $\Gamma \vdash M : T$ of SCI is a regular language.*

This justifies our choice of SCI as an interesting target language, as it has both higher-order functional features, imperative features, and a finite-state model at any term which recommends it for efficient hardware compilation.

> **Principle 3** *There is an immediate connection between the following game-semantic and hardware concepts:*
>
> - *An* occurrence *of a move is a* signal *on a port.*
> - *A* justified sequence *in an arena is a* waveform *on an interface.*
> - *The set of* plays *on an interface is a* protocol *of access to the interface.*

In the above we abstract the definition of a "signal," i.e. its precise phase encoding. Below, to keep the presentation focused we assume a standard low-high-low encoding, but other encodings can be equally used. This also affects the exact shape of the waveform.

Let us call the protocol corresponding to the SCI game model the *Sequentially Reentrant Access Protocol* (SRAP).

For instance, any communication mediated by interface ⟦com → com⟧ must behave following the following pattern:

1. the first action *must* be an input request on $q1$;

2. the second action *may* be an output request on $q2$ which indicates that the function is requesting an evaluation of its procedural argument

3. then, the third action *must* be an input acknowledgment on $a2$ indicating successful completion of the procedural argument's execution;

4. the protocol continues from Step 2;

5. the second action *may* also be an output acknowledgment on $a1$ indicating that the main procedure has completed execution.

If Steps 2 and 3 are omitted then we have a function that is ignoring its argument (i.e. non-strict). According to this protocol, all SRAP-compliant traces for this interface are captured by the regular expression $q1(q2 \cdot a2)^*a1$.

§

The game-semantic model is *denotational*, i.e. inductively defined on the syntax of the language. The model of a term is a *strategy*, i.e. a set of plays indicating how the Proponent will respond to any Opponent action, given the play up to that point. A strategy is subject to certain closure conditions, which are not relevant for our discussion here. More relevant is a property of strategies (*thread-independence* [11, Sec. 2.6.2]) that says that repeated plays of the same strategy will be essentially the same, i.e. the strategy has no hidden state information preserved between its initial question and its (final) answer. We call this property "*Reset*" noting that it is not a property of the protocol, but it is a property that circuits that participate in the protocol must satisfy. Note that it is known that circuits that are meant to be used compositionally must satisfy this self-resetting property [29].

One of the important properties of strategies is that they *compose* in a way that preserves both the legality of the resulting composite plays and the closure conditions of the resulting composite strategy. The details of composition are not relevant for our current discussion, but the following consequence is:

**Theorem 3 (Compositionality)** *If $A$ and $B$ are two SRAP-compliant circuits with interfaces ⟦$T_1 \to T_2$⟧ and ⟦$T_2 \to T_3$⟧ respectively, then the circuit $B \circ A$ of interface ⟦$T_1 \to T_3$⟧ formed by connecting the ports $T_2$ in the two circuits will be SRAP-compliant:*

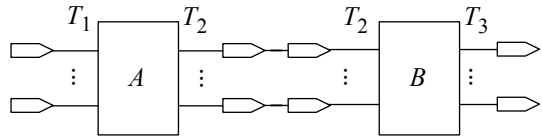

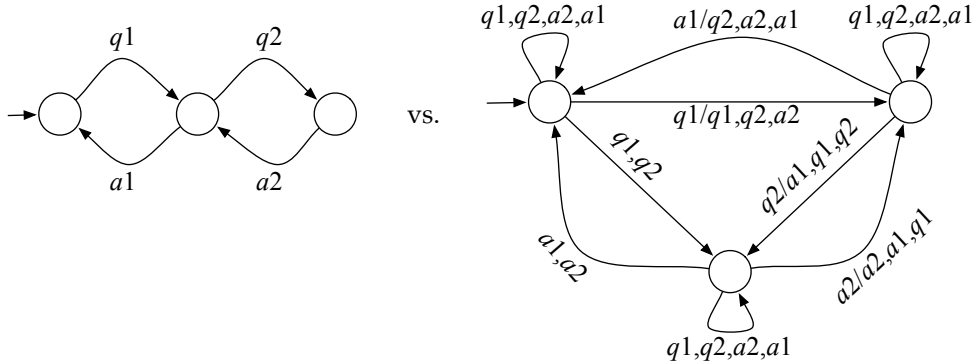

Figure 1: Asynchronous vs. synchronous model for `com` → `com`

The compositionality of a protocol rarely receives the emphasis it deserves. This property reduces substantially the burden of proof in creating complex composite circuits. It is enough to show that each individual module is protocol-compliant, then using compositionality we can safely assume that any interconnection of the sub-circuits that is compatible with the signature of the interface will result in a protocol-compliant circuit.

**Principle 4** *Interface-access protocols should be compositional in order to simplify the design of complex circuits.*

Note that the RESET property is also compositional in the same sense.

## 4.1 Synchronous versus asynchronous SRAP and related optimisations

Conventional game-semantic models exhibit an asynchronous view of concurrency. This makes them naturally suitable for the design of asynchronous hardware. When using game models to target synchronous (clocked) hardware one must take into account the possibility of several moves/signals occurring simultaneously. Note that this is level of synchronicity is simply used to to enhance performance, processing several inputs at once and issuing outputs instantaneously is faster, but has no bearing on correctness. A synchronous version of a protocol can be immediately derived using *round abstraction* [2]. For example, the (asynchronous) set of plays for `com` → `com` and its round-abstracted (synchronous) version can be seen in Fig. 1 (simultaneous signals are separated by commas, alternate labellings for a transition are separated by a slash).

The use of a asynchronous versus synchronous implementation has a serious impact on performance. The round-abstraction algorithm guarantees that the synchronous version of state machine has fewer (or equal) states, but may have more transitions. It is difficult to assess apriori the impact of round abstraction on the logical complexity of a circuit, but the synchronous version has an obvious speed advantage. Whereas the asynchronous model alternates between receiving inputs and producing outputs the synchronous model responds instantly (i.e. in the same clock cycle) to an input, and can handle several inputs at once. This makes implementation of constants and simple and ubiquitous functional constants (such as sequential composition) far more efficient, because they no longer introduce any delays at all, but respond instantly to input; in fact for both constants (true, false) and sequential composition the implementation is simply wires connected the right input and output ports.

The fact that our circuits work under a set protocol (SRAP) raises new opportunities for optimisation through automaton minimisation. Minimisation algorithms such as Hopcroft's [14] work by identifying bisimulation-equivalent states; however, when the protocol is set, it means that the automaton can only receive inputs from a restricted language, that of the protocol, which raises new possibilities for optimisation. Note that the more restrictive the protocol the greater the possibility of optimisation. The trivially extreme case of an empty protocol would allow any automaton to be reduced to the one-state automaton accepting all strings. Because we are working with a *fully abstract* game-semantic model, it means that the underlying protocol is the most restricted possible which still allows the compilation of any program. In other words, any finite interaction which is SRAP-compliant can be realised by a program [12].

This point is illustrate in Figs. 2–4. In it we show three basic programming language constructs: the `true` constant, sequential composition and iteration. In the figure we show the resulting configurations on an FPGA (Xilinx XC5VLX50) for the original asynchronous game model, the round-abstracted synchronous version and its optimisation under the assumed SRAP-compliant behaviour. These circuits are intended only to give a concrete feel for the effectiveness of the algorithms.

The reduction of `true` and `seq` to simply connectors is a remarkable optimisation which has a tremendous impact on the overall efficiency (time and space) of the circuits generated. Also, we should emphasise that the synchronous versions of the circuits we obtain are well-behaved under instant feedback, a well-known problem in conventional synchronous languages such as ESTEREL [4].

### 4.2 Activation managers

The development up to this point addresses gives a FIM consisting of a static definition (the interface) and a dynamic set of rules (the protocol) which solves the first problem we identified earlier: designing interfaces that correspond to (possibly higher-order) functions. In this section we shall see how re-instantiation can be avoided by using *sharing* instead. This also solves the problem of bookkeeping of names for modules and ports.

Suppose that we have a (slave) circuit $P$ with interface $[\![T]\!]$, and suppose it is used twice by a (master) circuit $Q$. In order to avoid instantiating $P$ twice we will use a special circuit called an *activation manager* (AM):

Asynchronous:

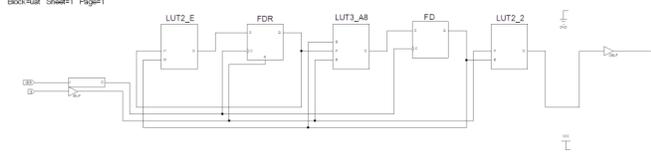

Synchronous:

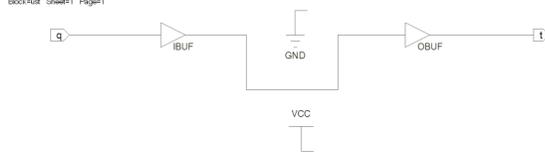

Minimised:

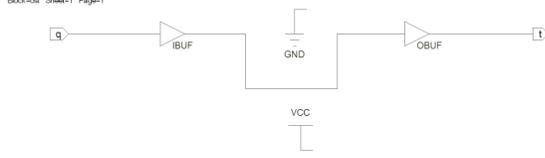

Figure 2: Circuit for the constant `true`

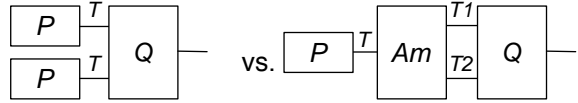

If $Q$ makes the initial request on a root port of $[\![T_1]\!]$ then AM will establish a logical connection between $P$ and the ports of $T_1$; it can be seen that AM works just like a local simple bus for $P$. If $P$ is large and AM small this can vastly improve the total footprint. We will see an example in the next section.

The question to ask at this point is: *How can we implement an AM?* Or even a more basic question: *What does it mean for an AM to be correct?*

Note that the AM is not one circuit, but a family of circuits indexed by any type in the language.

The correctness of the AM is precisely captured by the diagram above, which can be written equationally as

$$Q \circ \langle P, P \rangle = Q \circ Am \circ P.$$

In order for the composition to type-check, we need the circuits involved to have the following interfaces: $P : X \to T$, $Q : T \times T \to Y$ and $Am : T \to T \times T$. In fact the mention of $Q$ is irrelevant, what we need is simply

$$\langle P, P \rangle = Am \circ P.$$

These above are circuits, but now if we trace our intuitions back through the game-semantic model to the programming language it is quite obvious that we

Asynchronous:

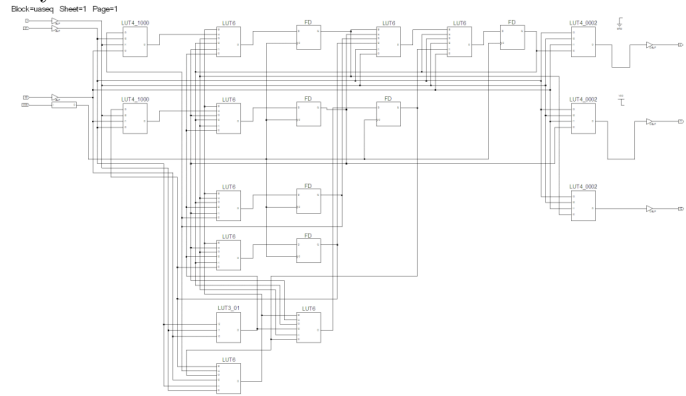

Synchronous:

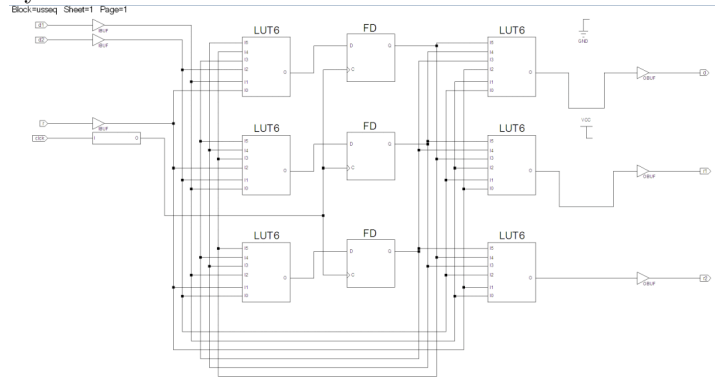

Minimised:

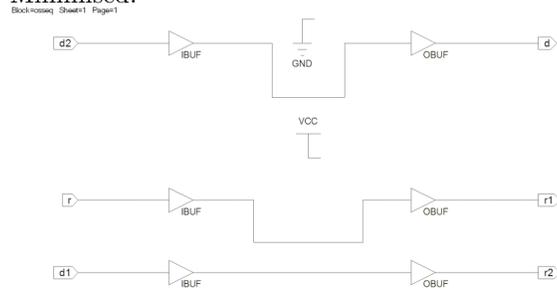

Figure 3: Circuit for sequential composition

Asynchronous:

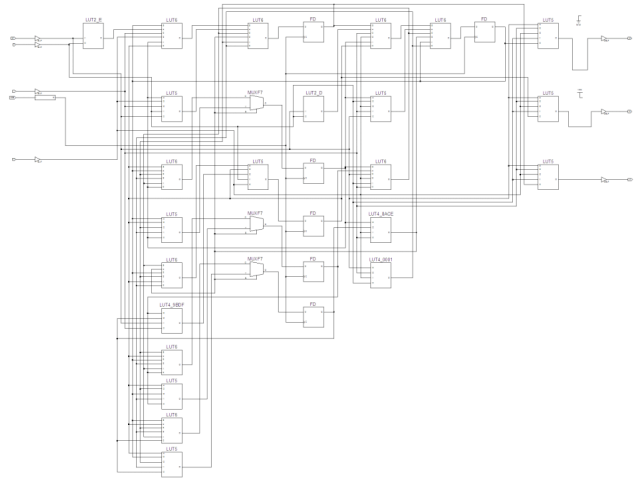

Synchronous:

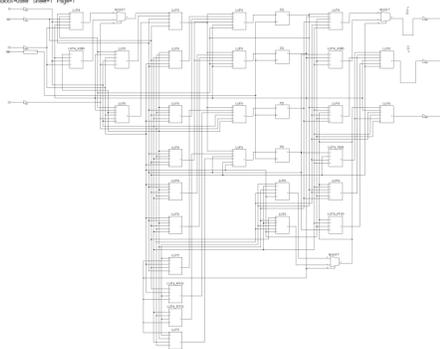

Minimised:

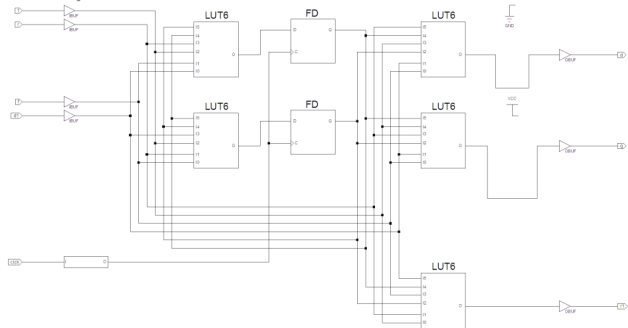

Figure 4: Circuit for iteration

want $Am$ to represent the denotation of the diagonal function $\lambda x.\langle x, x\rangle$ that given an argument returns a pairing of the argument with itself.

Note that this need for circuits to behave similarly under sharing versus replication is a main reason why they need to have the self-resetting property. Clearly, if $P$ above has internal state that is changes in a way that survives a round of usage—let us say that $P$ has an internal counter for the number of times it is activated—then it is not possible to expect the same behaviour in the shared scenario as in the replicated scenario, as the shared $P$ will be used twice as many times as the replicated instances.

Fortunately, we can extract a representation for the diagonal, at any type directly from the game-semantic model [12]: this is the Application Manager! From Thm. 2 we know that this is a finite-state model which can be mapped into hardware using standard methods (the synchronous version of the diagonal must undergo round abstraction, as explained before).

In Fig. 5 we show the asynchronous model for the diagonal $\texttt{com} \to \texttt{com} \times \texttt{com}$, the synchronous model and the circuits resulting from mapping the two diagonals on a Xilinx Virtex-5 (XC5VLX50) FPGA. The asynchronous AM requires 7 flip-flops (FF) and 22 look-up tables (LUT), while the synchronous version requires 3 FFs and 6 LUTs.

Finally, note that the AM for $\texttt{com} \to \texttt{com} \times \texttt{com}$ has both the same signature and the same behaviour as the CALL logical module in the *micro-pipelines* framework [28]. It is reasonable to claim that application managers are a generalisation across all type signatures of the concept of a CALL module.

### 4.3 Protocol compliance through type systems

As seen from the previous section, the proper operation of an AM is premised on it being used to connect SRAP-compliant, self-resetting components. The AMs themselves satisfy these conditions and, using the compositionality property (Thm. 3) the resulting circuits are also SRAP-compliant and can participate in other larger circuits structured using AMs. This is perfectly consistent with Harper's earlier quote:

> **Principle 5** *In hardware, type systems must be realised as compositional protocols.*

In hardware compilation the compositionality of SRAP simplifies the compilation process in the following way. Each basic construct of the language is compiled into a simple, SRAP-compliant circuit. Composition of sub-programs with disjoint sets of identifiers can be realised by wiring the corresponding circuits in a way that is consistent with the typing of their interfaces (cf. Theorem 3). Sharing of identifiers is then realised through AMs. The type system of the language will automatically allow sharing of identifiers in pairing but disallow it in function application. As already mentioned, according to the type system it is legal to form terms such as $\langle f(x); f(x)\rangle$ but illegal to form $f(f(x))$ or $f(x) \,||\, f(x)$.

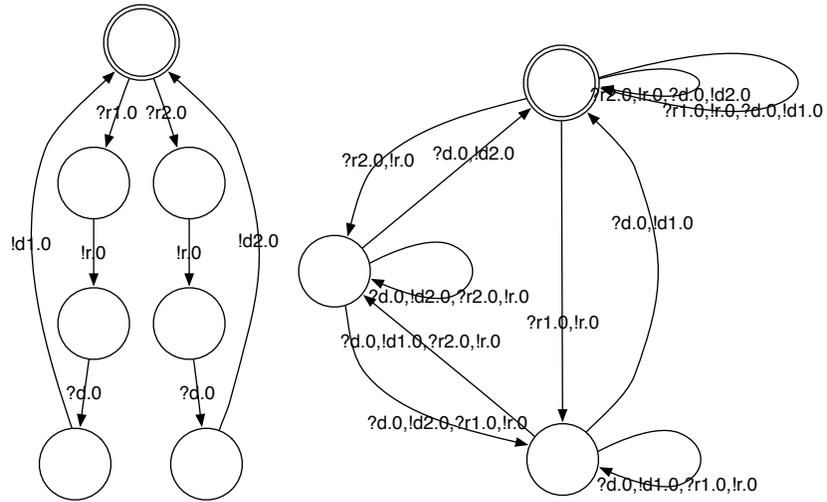
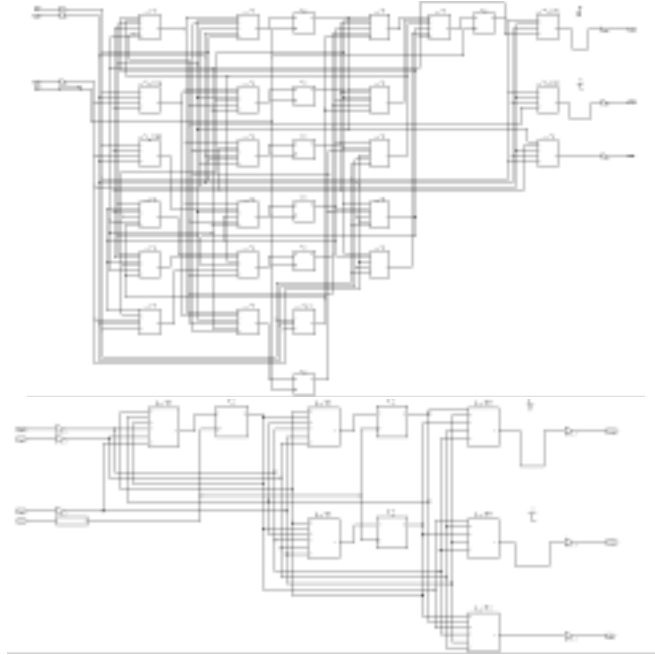

Figure 5: Asynchronous and synchronous AMs for $\mathtt{com} \to \mathtt{com} \times \mathtt{com}$

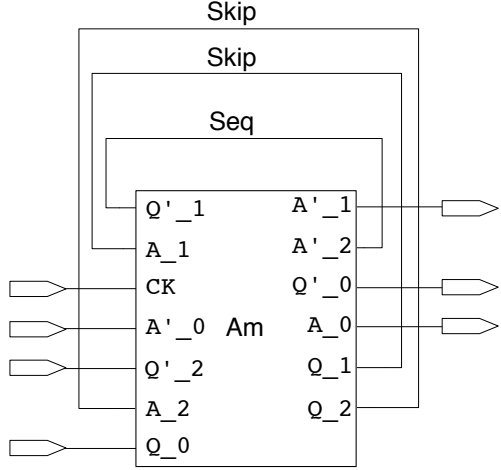

Figure 6: Sequential function call

Because the circuits are a direct representation of the semantics of the type system and the language, a compiler can be constructed denotationally. Function application is simply circuit composition, contraction is implemented by AMs of the right type, and lambda abstraction is, through the currying isomorphism, only a relabelling of ports. The implementations for the constants is a hardware representation of their game model. Additionally, the fact that all circuits are SRAP-compliant means that the FSM representation for their game model can be optimized beyond conventional minimisation [14], taking into account that certain input combinations, which violate the access protocol, are not possible [10].

## 4.4 Working with application managers

First lets look at an example of SRAP-compliant use of the activation manager. The representation of the game-semantic model (synchronous and optimized) of skip and seq are simply connectors, so the (legal) compiled program $f(\text{skip}); f(\text{skip})$ has schematic in Fig. 6, where AM is the diagonal for $(\text{com} \to \text{com}) \to (\text{com} \to \text{com}) \times (\text{com} \to \text{com})$.

Its behaviour is as follows: it will call the interface for $f$ twice, and each time $f$ asks for an argument it will provide skip, which is implemented as a wire between its own $Q$ and $A$ ports. A typical (legal) trace on this AM is

$$Q'_2 Q'_0 Q_0 Q_2 A_2 A_0 A'_0 A'_2 Q'_1 Q'_0 Q_0 Q_1 A_1 A_0 A'_0 A'_1.$$

If we attempt to synthesise the invalid program $f(f(\text{skip}))$ we obtain the schematic in Fig. 7, with skip implemented as a connector and function application implemented as wiring.

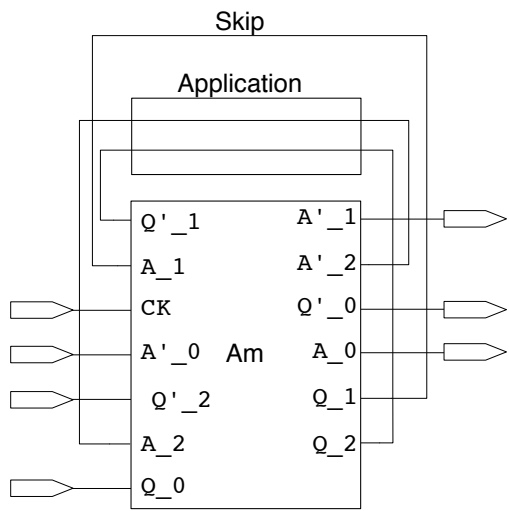

Figure 7: Nested function call

To be more precise, the circuit above is

$$\lambda f.(\lambda f'.(\pi_1 f')((\pi_2 f')(\text{skip})))(\delta f)).$$

Where $\delta = \lambda x.\langle x, x \rangle$ is the diagonal, implemented as $Am$.

The AM, same as before, must handle now traces of the form

$$Q'_2 Q'_0 Q_0 Q_2 Q'_1 Q'_0 Q_0 Q_1 A_1 A_0 A'_0 A'_1 A_2 A_0$$

Note the nesting of request and acknowledgements $Q_0$ and $A_0$, which breaks serialization. Simulating the circuit in a tool such as Xilinx ISE we can see that the activation manager simply deadlocks after this trace and it never produces the $A_0$. The reason is that after $A'_1$ the AM has established a communication channel with the first projection, so the input on $A_2$ is unexpected and cannot be handled. Keeping track of nested function calls requires more than static resources and is substantially more expensive to implement.

§

The second typical way to violate SRAP is by trying to use concurrent access to interface components, as in the the program for $\text{par}(f(\text{skip}))(f(\text{skip}))$, synthesised as in Fig. 8, where par is part of the implementation of parallel composition. The problems here are obvious and no trace analysis is required. AM is not equipped to arbitrate the simultaneous initial requests on $Q'_1, Q'_2$, but even if it did, they are forwarded *at the same time* to the *same* instance of $f$ via $Q'_0$! This is precisely the source of the bug in HANDELC.

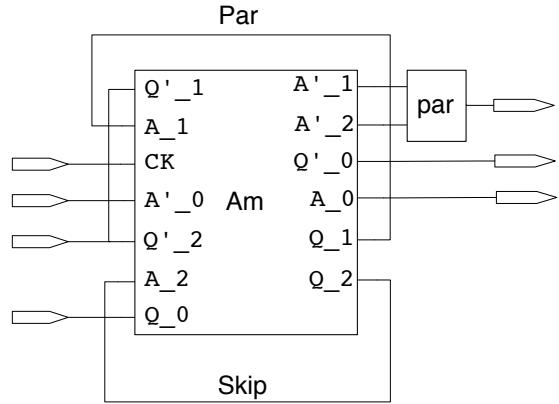

Figure 8: Concurrent function call

## 5 Conclusion

The effectiveness of this approach can be easily argued, as all the weaknesses enumerated before, arising form the lack of a FIM are resolved using the games-based approach. We are able to define interfaces that implement higher-order functions and thanks to the well-defined SRA protocol we can construct (automatically) light-weight AMs, at least as compared to alternatives such as buses or NOC, that can efficiently (in time and space) share (self-resetting) modules with any SRAP-compliant signature.

An earlier version of this work was the *Geometry of Synthesis* (GOS) [9], a direct circuit-based semantics for a similar language (lacking parallel composition). The difference between GOS and the simpler method we propose here is that the former is inspired by game semantics, whereas the latter is simply game semantics. Using the game semantic model has certain advantages, such as eliminating the need for re-proving soundness. Using the game-semantic model also has some other, subtler advantages that take advantage of the full-abstraction property of the model to realize further optimizations of the concrete representation of the game model [10].

Also, establishing a methodology for the semantic-directed compilation of game models into hardware, preserving correctness, allows us to take advantage of the rich game-semantic literature, which provides fully abstract models for a large variety of programming languages.

A proof-of-concept compiler was implemented using this approach. One of the test applications was a back-propagation neural-network with 12 4-bit neurons. The project was carried out by a student with no training in digital design. Using a standard textbook algorithm of 167 lines of source code, the compiler produced a design that fitted comfortably on a Spartan XC3S200 FPGA, using 10% of available FFs, 13% of available LUTs and 19% of available slices, clocked at maximum frequency.

The argument of this paper is a methodological one: FIMs, access protocols, application managers and language-support through type systems, all fit naturally in the game-semantic framework and can be immediately and naturally translated into relevant hardware concepts. These concepts can overcome the absence of FIMs in hardware compilation, which is a significant obstacle in the way of producing mature and usable hardware compilers. Without well defined FIMs it is not possible to support separate compilation units, which greatly restricts library support and run-time inter-operability of components. Another immediate consequence of having a well defined FIM is the ability to implement activation managers, circuits that can share the functionality of other circuits. We have presented a simple (but non-trivial) example for a FIM and an associated access protocol that guarantees the correct usage of circuits that implement the interface. The SRA Protocol supports a rich type system and has built-in support for safe concurrency, which is an important step forward compared to the state-of-the-art in hardware compilation.

**Principle 6** *Lessons learned decades ago concerning the design and implementation of compilers must not be ignored, but must be adapted to hardware compilation.*